# Characterising Solar Magnetic Reconnection in Confined and Eruptive Flares

Kanniah BALAMURALIKRISHNA[1], John Y. H. SOO[1], Norhaslinda MOHAMED TAHRIN[1] and Abdul Halim ABDUL AZIZ[1]

[1] School of Physics, Universiti Sains Malaysia, 11800 USM, Pulau Pinang, Malaysia.

**Abstract**

Magnetic reconnection is a fundamental mechanism through which energy stored in magnetic fields is released explosively on a massive scale, they could be presented as eruptive or confined flares, depending on their association with coronal mass ejections (CMEs). Several previous works have concluded that there is no correlation between flare duration and flare class, however, their sample sizes are skewed towards B and C classes; they hardly represent the higher classes. Therefore, we studied a sample without extreme events in order to determine the correlation between flare duration and flare type (confined and eruptive). We examined 33 flares with classes between M5 to X5 within 45° of the disk centres, using data from the Atmospheric Imaging Assembly (AIA) and the Helioseismic and Magnetic Imager (HMI). We find that the linear correlation between flare class against flare duration by full width half maximum (FWHM) in general is weak ($r = 0.19$); however, confined flares have a significant correlation ($r = 0.58$) compared to eruptive types ($r = 0.08$). Also, the confined M class flares' average duration is less than half of the eruptive flares. Similarly, confined flares have a higher correlation ($r = 0.89$) than eruptive flares ($r = 0.60$) between flare classes against magnetic reconnection flux. In this work, a balanced sample size between flare types is an important strategy for obtaining a reliable quantitative comparison.

## 1 Introduction

Solar flares are sudden electromagnetic radiation associated with magnetic reconnection, which is a fundamental mechanism through which energy stored in magnetic fields is released explosively on a massive scale (Wagner, 1984), this involves the restructuring of complex magnetic topologies within the corona (Aarnio et al., 2011). Their magnetic configuration rearrangement process transforms the magnetic energy to kinetic and thermal energy, causing the acceleration of non-thermal particles (Priest & Forbes, 2002; Shibata & Magara, 2011).The rapid conversion of free magnetic energy stored in the sheared and twisted magnetic fields of active regions leads to solar flare emissions across the whole range of electromagnetic wavelengths (Fletcher et al., 2011; Forbes, 2000; Hudson, 2011; Kazachenko et al., 2012). The coronal mass ejection (CME) particles could reach up to $10^{12}$ kg at speeds ranging from hundreds to a few thousand kilometres per second, with an average value of ∼ 490 km s$^{-1}$ (Gosling et al., 1976; Hundhausen et al., 1994; Gopalswamy et al., 2001; Gallagher et al., 2003; Webb & Howard, 2012). Solar flares and CMEs are significant phenomena that need further investigation and understanding. Therefore, it is important to review the observations and physical mechanisms behind solar eruptions, flares, and CMEs and to assess the current capability of forecasting these events for space weather risk and impact mitigation (Green et al., 2018).

CMEs can occur under two models, known as the ideal instability model and resistive model. The ideal instability model is based on the destabilisation of magnetic field configurations without invoking magnetic reconnection. This model explores plasma instabilities, such as the kink and torus instability. The kink instability model for CMEs was introduced by Kliem & Török (2006), which is based on the concept that a twisted magnetic flux rope can become unstable and undergo a kink instability, leading to the ejection of plasma and magnetic fields into interplanetary space. The torus instability model for CMEs was proposed by Forbes & Priest (1984). This model is based on the idea that a magnetic flux rope can become unstable and undergo a torus instability, leading to the ejection of plasma and magnetic fields into interplanetary space. Both models are based on ideal instabilities, which provide quantitative onset criteria that can be applied to observations or magnetic-field extrapolations of potential CME source regions. However, the exact thresholds of these instabilities depend on the detailed structure of the magnetic-field configuration, which is not sufficiently well known.

Another model is the resistive model, which incorporates magnetic reconnection. Resistive models that invoke magnetic reconnection for CMEs have been proposed by several authors. Antiochos et al. (1999) proposed the breakout model for CMEs, which involves magnetic reconnection. In this model, a current sheet forms between the magnetic fields of two adjacent flux systems, and magnetic reconnection occurs, leading to the release of magnetic energy and the ejection of plasma into interplanetary space. Lynch et al. (2008) developed a resistive magnetohydrodynamic (MHD) model for CMEs that includes magnetic reconnection. In this model, the magnetic field in the corona becomes highly stressed and distorted, leading to the formation of a current sheet and magnetic reconnection. This can result in the release of magnetic energy and the ejection of plasma into interplanetary space. More recently, Manchester et al. (2017) used a resistive MHD model to simulate the evolution of a CME from its onset in the low corona out to 1 AU. Their model includes magnetic reconnection, and they found that the CME is driven by a combination of magnetic and thermal pressure gradients, along with magnetic reconnection. Magnetic reconnection is a central process driving CMEs in this model, the magnetic field lines undergo rearrangement, allowing for the release of stored magnetic energy, followed by the eruption of plasma as a CME, which is referred to as an eruptive event in this study. Otherwise, it is called confined events without CME association (Švestka & Cliver, 1992).

CMEs and flares are considered different consequences of coronal magnetic field instability and reconnection (Lin et al., 2003; Forbes et al., 2006; Wiegelmann et al., 2014). CMEs may generally occur without flares (Robbrecht et al., 2009; D'Huys et al., 2014), and flares may occur without CMEs (Hudson, 2011; Sun et al., 2015). However, the association rate is a steeply increasing function of flare class, and both typically occur together in the strongest flare events (Andrews, 2003). A careful statistical study on the CME-flare association was carried out by Yashiro et al. (2006); the results show that the association rate of CMEs with flares is about 50% for M class flares, while the percentage increases up to > 90% for X1.0 flares and almost 100% for X2.0 flares and above.

Solar flare essential parameters such as their size as determined by the peak X-ray flux, flare duration, and appearance rate are continuously monitored by the Geostationary Operational Environmental Satellite (GOES). The flare classification is based on the peak magnitude of the 1–8 Å band, generally accepted as a standard measurement of the flare's magnitudes. Their characteristics have been extensively studied using GOES soft X-ray (SXR) critical parameters. For example, Toriumi et al. (2017) analysed 51 flares greater than M5 and discovered a strong linear correlation between flare duration and ribbon area, total flux, and ribbon separation distance, with coefficient values of 0.72, 0.79, and 0.83, respectively.

In another related study conducted by Kazachenko et al. (2017), a larger sample size of 3137 flare events with a class C1.0 and greater was analysed. The result revealed a significant correlation between the flare class and both the cumulative flare ribbon area and the flare ribbon reconnection flux, with correlation coefficients of 0.68 and 0.66, respectively. This agrees with recent findings made by Li et al. (2020), where a dataset of 322 solar flares ranging from class M1.0 and greater, consisting of 152 confined events and 170 eruptive events was analysed. Their study found that the amount of magnetic flux in active regions (ARs) is a crucial factor in determining the type of flare. Flares originating from ARs with high magnetic flux are mostly confined, while those from ARs with low magnetic flux are mostly eruptive. The study also revealed that ARs with high magnetic flux have a strong magnetic cage that confines the eruption, which is a novel finding. Li et al. (2020) observed a moderate correlation between the peak X-ray flux and the flare ribbon reconnection flux for all flares, with a correlation coefficient of 0.51. The correlation coefficient increased to 0.58 for eruptive flares and decreased to 0.42 for confined flares. Similarly, the correlation coefficient between peak X-ray flux and flare ribbon area was 0.58 for all flares, 0.48 for eruptive flares, and 0.53 for confined flares. These results are consistent with the findings of Veronig & Polanec (2015), Kazachenko et al. (2017), and Tschernitz et al. (2018).

In addition, Reep and Toriumi (2017) established a direct correlation between the duration of magnetic reconnection and the duration of the GOES light curves due to the increasing length of the loops that span the ribbon separation. In another study on flare duration, Zhang and Liu (2015) observed that the rise time of the soft X-ray flux of a flare is approximately half of the decay time, and the rise and decay times increase with the increasing flare class magnitude.

Recent studies by Reep and Knizhnik (2019) and Veronig et al. (2002) show that flare duration is independent and weak, respectively, against flare class. However, there are no conclusive results on flare duration compared to the type of flare. Additionally, the sample size was skewed towards certain lower flare classes (B and C-classes) and thus could not accurately reflect the higher magnitude X-class flares. Harra et al. (2016) showed that X-class flare duration has no correlation with flare class and type, however, their sample size is skewed towards eruptive flares due to the limited occurrence of confined X-class events due to CME usually associated with larger flares (Yashiro & Gopalswamy, 2009; Youssef, 2012). In another study, Veronig & Polanec (2015) discovered a significant correlation between the total magnetic reconnection flux against flare class, while Tschernitz et al. (2018) found a slightly higher correlation coefficient for eruptive flares as compared to confined flares. However, in the population sample size, there were extreme events included that could be regarded as outliers (Miklenic et al., 2009), and additionally, converting the variables to a logarithmic scale sometime could yield a higher correlation value as in this case.

The vital aspect of the investigation in this study is to analyse the correlation of total magnetic reconnection flux and flare duration (FHWM) against flare classes while comparing confined and eruptive flares separately. Due to this reason, the sample selection for both confined and eruptive flare types only consisted of comparable sample size flares between M5 and X5 and avoided any (few) extreme flares such as X10 or X17 from affecting our results. We hope that this effort will uncover previously unexplored portions of earlier data and help future research determine the sample size for solar flare-type events.

Therefore, correlation coefficients are derived using a balanced sample size by flare types (confined and eruptive) to:

1. Determine the correlation coefficient of *flare class against duration*, before and after segregation by flare type; and
2. Determine the correlation coefficient of *flare class against total magnetic reconnection flux*, before and after segregation by flare type.

This unbiased comparison helps to reflect the relationship between these variables quantitatively and more specifically according to flare type, rather than a common conclusion representing all flare types.

## 2 Methodology

The data used in this work are taken from the sources described in the following sections. In the following paragraphs, we will explain the methodology used to derive the flare duration and total magnetic reconnection flux for each of the solar flare events used in this work.

### 2.1 Solar Flare Event Selection

Solar flare events are classified based on the peak magnitude value of their X-ray flux as measured by the GOES X-ray flux sensor. The flare magnitudes corresponding to GOES classes between M5 to X5 were initially identified. The events were then filtered based on their location within heliographic longitudes less than 45° using online resources like Solar Monitor (https://solarmonitor.org) and Space Weather (https://www.spaceweatherlive.com). Subsequently, the events were shortlisted by confined and eruptive flare type, which was determined through literature references (Harra et al., 2016; Toriumi et al., 2017; Mitra et al., 2020). Thus, we examined the magnetic reconnection fluxes and flare durations (represented by the full width at half maximum, FWHM) of 33 flare events, consisting of 17 confined and 16 eruptive flares. The events are dated from February 2011 to September 2017, covering the entirety of Solar Cycle 24. As shown in **Table 1**, the information compiled includes the dates, the National Oceanic and Atmospheric Administration (NOAA) active region number, location, CME status, flare type, and event reference.

### 2.2 Flare Duration by FWHM in GOES X-Ray

Time series data from the NOAA GOES X-ray flux archives (https://satdat.ngdc.noaa.gov/sem/goes/data/full/) are fundamental to solar tracking and analysis, therefore they were used to derive the FWHM. The data for the long wavelength channel (1–8 Å) are available by date with a tracking interval of 2 s, they are used for flare classification and duration. For each event, we narrow down the range of records near the X-ray flux peak value and obtain sufficient data to cover the beginning and the end of the solar flare. By analysing the X-ray intensities (B_FLUX, in $Wm^{-2}$) and the corresponding time series (TIME_TAG), we determine the peak intensities of the flares and derive the flare durations as the FWHM, which measures the duration between the two points $t_2 - t_1$, where the curve intersects the half maximum value during the rise and decay phase. An example of these data for a flare that occurred on 28 Sep 2015 is illustrated in **Figure 1**.

**Table 1**: The list of solar flare events used in this work. Events 1-17 are confined (no CME), while events 18-33 are eruptive (with CME).

| No. | Event (Date) | GOES Class | NOAA Number | Position | Start Time | Peak Time | End Time | Flare Type | Event Reference |
|---|---|---|---|---|---|---|---|---|---|
| 1 | 3-Nov-13 | M5.0 | 11884 | S12W13 | 5:16 | 5:22 | 5:26 | Confined | Toriumi et al., 2017 |
| 2 | 4-Feb-14 | M5.2 | 11967 | S13W12 | 3:57 | 4:00 | 4:06 | Confined | Toriumi et al., 2017 |
| 3 | 4-Jul-12 | M5.3 | 11515 | S16W10 | 9:47 | 9:55 | 9:57 | Confined | Toriumi et al., 2017 |
| 4 | 24-Aug-15 | M5.6 | 12403 | S14E00 | 7:26 | 7:33 | 7:35 | Confined | Toriumi et al., 2017 |
| 5 | 10-May-12 | M5.7 | 11476 | N10E22 | 4:11 | 4:18 | 4:23 | Confined | Toriumi et al., 2017 |
| 6 | 5-Jul-12 | M6.1 | 11515 | S17W23 | 11:39 | 11:44 | 11:49 | Confined | Toriumi et al., 2017 |
| 7 | 13-Feb-11 | M6.6 | 11158 | S19E11 | 17:28 | 17:38 | 17:47 | Confined | Toriumi et al., 2017 |
| 8 | 18-Dec-14 | M6.9 | 12241 | S10E19 | 21:41 | 21:58 | 22:25 | Confined | Toriumi et al., 2017 |
| 9 | 7-Jan-14 | M7.2 | 11944 | S09E11 | 10:07 | 10:13 | 10:37 | Confined | Toriumi et al., 2017 |
| 10 | 28-Sep-15 | M7.6 | 12422 | S20W16 | 14:53 | 14:58 | 15:03 | Confined | Toriumi et al., 2017 |
| 11 | 22-Oct-14 | M8.7 | 12192 | S14E19 | 1:44 | 1:59 | 2:28 | Confined | Toriumi et al., 2017 |
| 12 | 30-Jul-11 | M9.3 | 11261 | N14E35 | 2:04 | 2:09 | 2:12 | Confined | Toriumi et al., 2017 |
| 13 | 19-Oct-14 | X1.1 | 12192 | S13E42 | 4:17 | 5:03 | 5:48 | Confined | Harra et al., 2016 |
| 14 | 9-Mar-11 | X1.5 | 11166 | N11W01 | 23:13 | 23:23 | 23:29 | Confined | Harra et al., 2016 |
| 15 | 22-Oct-14 | X1.6 | 12192 | S14E19 | 14:02 | 14:28 | 14:50 | Confined | Harra et al., 2016 |
| 16 | 26-Oct-14 | X2.0 | 12192 | S12W34 | 10:04 | 10:56 | 11:18 | Confined | Harra et al., 2016 |
| 17 | 24-Oct-14 | X3.1 | 12192 | S14W06 | 21:07 | 21:41 | 22:13 | Confined | Harra et al., 2016 |
| 18 | 10-Mar-15 | M5.1 | 12297 | S15E39 | 3:19 | 3:24 | 3:28 | Eruptive | Toriumi et al., 2017 |
| 19 | 28-Sep-14 | M5.1 | 12173 | S13W23 | 2:39 | 2:58 | 3:19 | Eruptive | Toriumi et al., 2017 |
| 20 | 6-Sep-11 | M5.3 | 11283 | N14W04 | 1:35 | 1:50 | 2:05 | Eruptive | Toriumi et al., 2017 |
| 21 | 4-Dec-14 | M6.1 | 12222 | S20W22 | 18:05 | 18:25 | 18:56 | Eruptive | Toriumi et al., 2017 |
| 22 | 9-Mar-12 | M6.3 | 11429 | N17E01 | 3:22 | 3:53 | 4:18 | Eruptive | Toriumi et al., 2017 |
| 23 | 22-Jun-15 | M6.6 | 12371 | N13E14 | 17:39 | 18:23 | 18:51 | Eruptive | Toriumi et al., 2017 |
| 24 | 8-Sep-11 | M6.7 | 11283 | N14W32 | 15:32 | 15:46 | 15:52 | Eruptive | Romano et al., 2015 |
| 25 | 7-Sep-17 | M7.3 | 12673 | S10W43 | 10:11 | 10:15 | 10:18 | Eruptive | Mitra et al., 2020 |
| 26 | 25-Jun-15 | M7.9 | 12371 | N11W40 | 8:02 | 8:16 | 9:05 | Eruptive | Toriumi et al., 2017 |
| 27 | 17-Dec-14 | M8.7 | 12242 | S18E08 | 4:25 | 4:51 | 5:20 | Eruptive | Toriumi et al., 2017 |
| 28 | 4-Aug-11 | M9.3 | 11261 | N16W37 | 3:41 | 3:57 | 4:04 | Eruptive | Toriumi et al., 2017 |
| 29 | 8-Nov-13 | X1.1 | 11890 | S13E13 | 4:20 | 4:26 | 4:29 | Eruptive | Harra et al., 2016 |
| 30 | 12-Jul-12 | X1.4 | 11520 | S17E06 | 15:37 | 16:49 | 17:30 | Eruptive | Harra et al., 2016 |
| 31 | 10-Sep-14 | X1.6 | 12158 | N11E05 | 17:21 | 17:45 | 18:20 | Eruptive | Harra et al., 2016 |
| 32 | 11-Mar-15 | X2.2 | 12297 | S17E22 | 16:11 | 16:22 | 16:29 | Eruptive | Toriumi et al., 2017 |
| 33 | 7-Mar-12 | X5.4 | 11429 | N17E29 | 0:02 | 0:24 | 0:40 | Eruptive | Harra et al., 2016 |

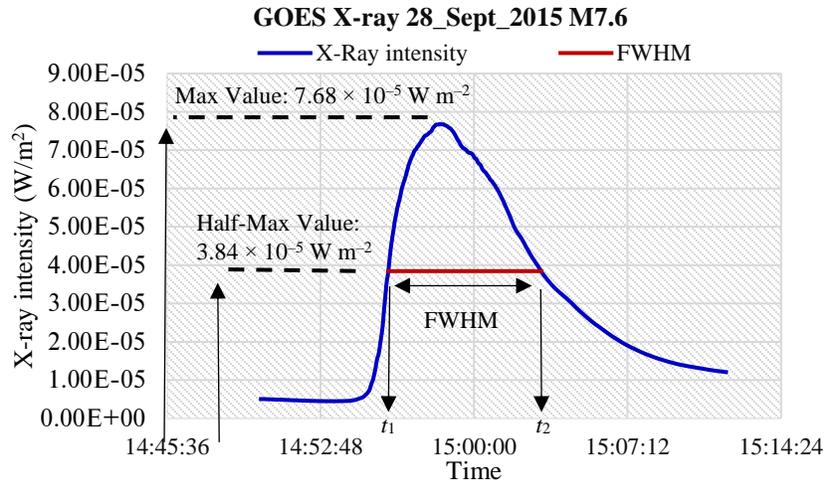

**Figure 1**: FWHM derivation from GOES X-ray flux data of a flare that occurred on 28 Sep 2015.

## 2.3 Total Magnetic Reconnection Flux

Magnetic reconnection releases kinetic and thermal energy into the plasma, which causes particle acceleration. The chromosphere flare ribbon evolution images indicate the newly reconnected magnetic field. In line with Forbes & Priest (1984), Veronig & Polanec (2015), and Kazachenko et al. (2022), we use the newly brightened area of an image with an underlying magnetic field as a proxy for the total magnetic reconnection flux. The total magnetic reconnection flux cannot be derived directly but instead has to be obtained via an indirect approach. The magnetic reconnection occurs in the corona but is detected on the chromosphere, while the magnetic properties are compiled from the photosphere; therefore, archived filtergram images have to be used to identify the flare-ribbons and magnetic fluxes.

### *2.3.1 Data Preparation*

The filtergram 1600 Å by the Atmospheric Imaging Assembly (AIA, Lemen et al., 2012) has been used in this study to identify the flare-ribbon area, since it is the most sensitive filter in capturing radiation at 10 000 K, the temperature of the region between the upper chromosphere and the transition region. The magnetic flux data are obtained from the Helioseismic and Magnetic Imager (HMI, Schou et al., 2012), magnetogram maps of the solar magnetic field in the photosphere at 6173 Å. Both instruments are aboard the Solar Dynamics Observatory (SDO) satellite, whose data records are accessible through the Joint Science Operations Centre (JSOC, http://jsoc.stanford.edu). Here we note that the SDO images from AIA (1600 Å) and HMI line-of-sight magnetic field maps were taken with cadences of 24 s and 45 s, respectively; thus in pairing the images, a discrepancy of interval time 0–12 s is unavoidable.

The AIA images were rescreened to identify any instances of blooming defects, which occur when the charges of saturated pixels spill over into adjacent pixels. These blooming defects can result in inaccurate flare-ribbon coverage. Hence, manual inspection of all AIA images was performed using the data visualisation tool SAOImageDS9 (DS9, Joye & Mandel, 2003), and detected blooming defect frames were omitted from further analysis. In future studies, the criterion for removing saturated pixels, as proposed by Veronig & Polanec (2015) and Thalmann et al. (2015), shall be applied. This analytical approach is more systematic and has the potential to lessen the discrepancy caused by the blooming defects. Examples of blooming defects is displayed in **Figure 2**.

The pixel intensity in an AIA image can spike up to a maximum value of about 18 000 units during magnetic reconnection; therefore, a standard limit is needed to differentiate between pre-flare and flare pixels. A conventional approach is used to calculate the threshold, with a fixed cut-off value of 10% higher than the maximum intensity value during pre-flare as a background. Thus, a pixel value greater than the threshold of 1300 units is counted as a flare-ribbon pixel in this study, which is comparable to the value of 1590 used by Veronig & Polanec (2015). Their threshold value was accomplished by using image analysis software, which adaptively determines the threshold value depending on the brightness histogram of the image (Qiu et al., 2010). Its algorithm seeks a threshold that distinguishes between foreground and background objects while minimizing the variation between the two classes. In a conventional approach, we used the astronomical imaging and data visualisation application DS9. Generally, (Miklenic et al., 2009a) tested several threshold values within a reasonable range and found that the magnetic reconnection rate may result in some variation from 5 to 15%.

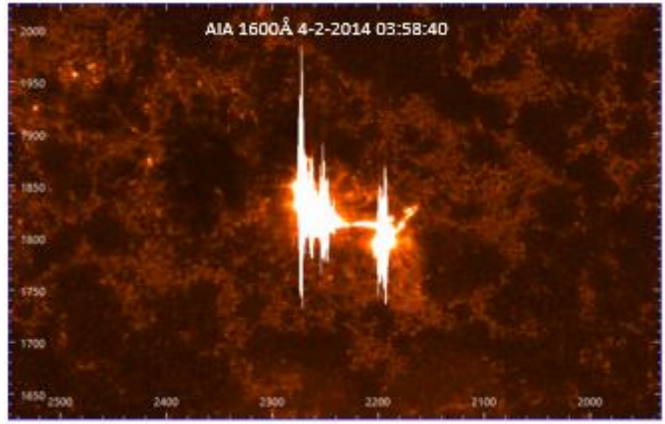

**Figure 2**: The image AIA 4-2-2014 showed an example of blooming defect.

## 2.3.2 Determination of the Total Magnetic Reconnection Flux

Since the AIA and HMI images were taken from different instruments (one telescope and the other a doppler camera), the images are in different orientations by default. The two images are fitted together by aligning their geometric differences for each pixel (co-registering), and later superimposed over each other for each observation time (masking). The flare-ribbon shown in the HMI image masked under the AIA image is the magnetic parameter required for us to measure the total magnetic reconnection flux. The newly brightened pixels are only counted during masking in every subsequent frame.

The image processing task has been accomplished through a software package developed using the python library SUNPY. It is a community-developed free and open-source software package specially designed for solar physics (The SunPy Community et al., 2015). The total magnetic reconnection fluxes were obtained for all the events using our algorithm, working on each event's pairing AIA and HMI images as inputs for image processing. The summary of the code sequence is as follows:

1. For each event, the AIA image frame with the highest number of pixels with intensities > 1300 (the threshold) is identified to set the area of interest (rectangular frame). This frame covers the area of interest to be analysed starting from the initial frame of the event.
2. In the selected AIA image frame, the coordinates of the pixel with the highest intensity are used as the centre point to create a rectangular box with length $600 \times 600$ pixels, which will act as the perimeter that covers the flare ribbon area for all AIA images of the same event. The image frames are processed as follows:
   - The centre point coordinate of the rectangular box frame is referred as (0,0);
   - The box width ($w$) and height ($h$) are recorded and used to determine the coordinates of the bottom-left ($x_{bl}$, $y_{bl}$) and top-right ($x_{tr}$, $y_{tr}$) corner of the box,
   $$(x_{bl}, y_{bl}) = \left(-\frac{w}{2}, -\frac{h}{2}\right), \qquad (x_{tr}, y_{tr}) = \left(\frac{w}{2}, \frac{h}{2}\right);$$
   - The SUNPY function `pixel_to_world` converts the pixel coordinates to world coordinate units;
   - A submap is created from the original AIA map using the calculated bottom-left and top-right coordinates. The AIA submap is used to define the HMI submap, extracting a smaller region of interest from the original map based on the specified coordinates;

- Both submaps are rotated to be aligned with the north direction using the SUNPY function `rotate` from the `Map` class, with a rotation matrix calculated based on the submap's properties.
3. Starting from the first AIA image frame, the pixels with intensities > 1300 (known as 'brightened pixels' from this point onwards) within the perimeter box are identified. In the second image frame, only the number of newly brightened pixels are counted (i.e. the pixels that had intensities < 1300 in the first frame but have intensities > 1300 in the second frame), while pixels which remain brightened are omitted. This procedure is repeated for every subsequent consecutive image frame, until the last frame.
4. The AIA image frames are co-registered with their respective HMI image frames of the same timestamp, and the locations of the newly brightened pixels are masked over from the AIA to HMI image frames. The mask is updated by removing the AIA pixels that have already been processed to ensures that each pixel is considered only once. The corresponding pixels in the HMI image are extracted, along with their magnetic values.
5. The newly brightened pixels now on the HMI image frames are identified as the magnetic reconnection pixels. The magnetic flux densities are summed up separately according to their polarity and recorded for further processing.

The output of the algorithm contains the analysed image frames, coordinates, and magnetic flux values. The HMI measures the magnetic field strength of the Sun in Gauss (G), which is then converted into Maxwell (Mx) units as it is more commonly used in recent publications (1 Mx = G cm$^2$). For each event, we plotted the cumulative sum of the magnetic flux values separated by polarity against time along with the corresponding GOES X-ray flux time series; an example is visualised in **Figure 3**. A newly brightened pixel is a proxy of energy released by a magnetic reconnection flux; thus, accumulation of total magnetic reconnection flux represents the total magnetic reconnection energy released respective to the polarity and their mean value represents the total magnetic reconnection flux of the event.

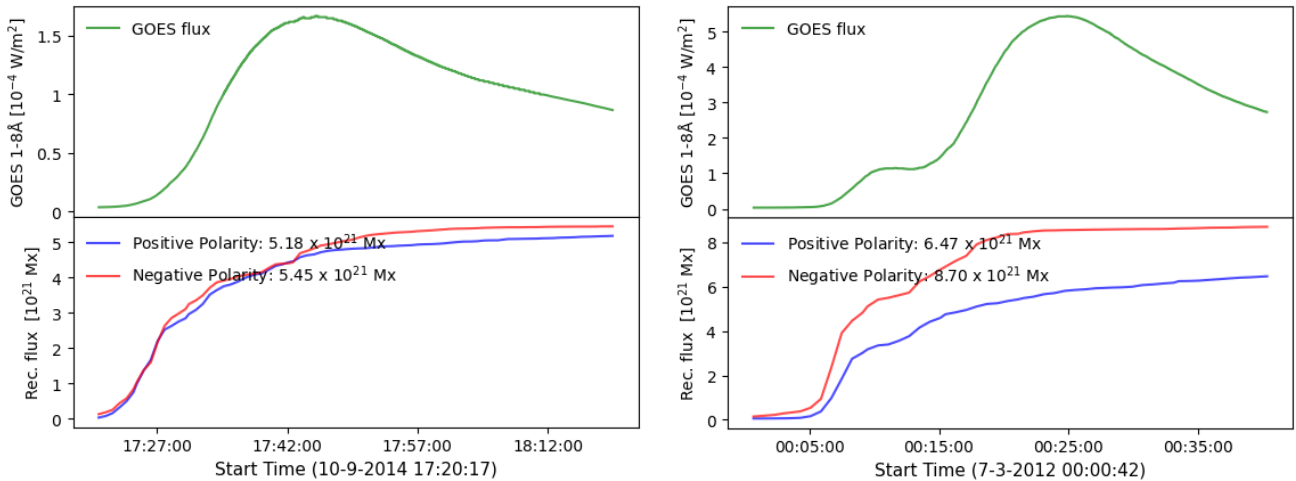

**Figure 3**: GOES 1 – 8 Å soft X-ray flux trend (top) and accumulation magnetic reconnection flux (bottom) by polarity trends.

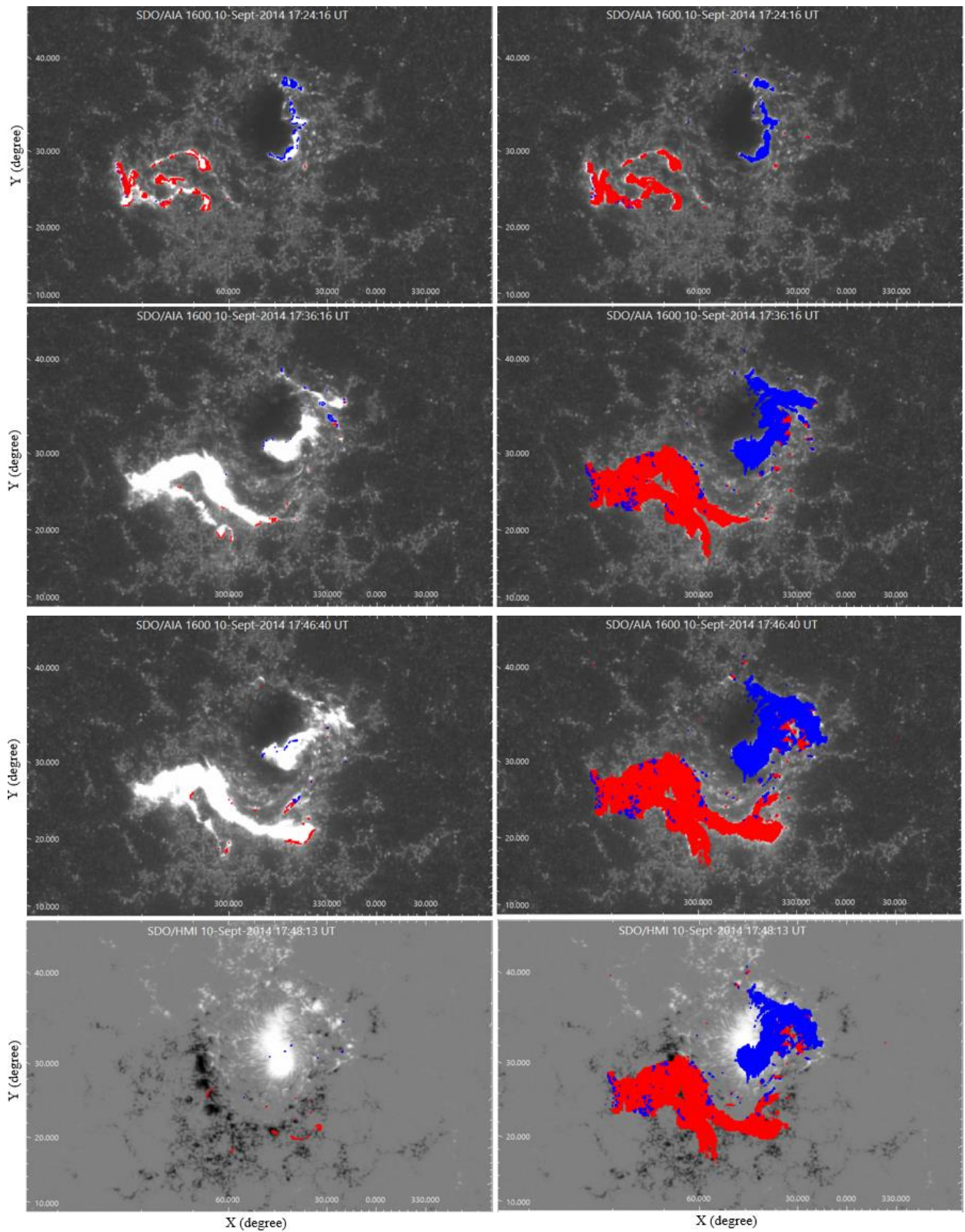

**Figure 4**: The interval image frames for Event X1.6 of 10-9-2014, where the initial 3 rows display SDO/AIA images of the flare ribbon evolution, while the bottom row shows the SDO/HMI magnetograms according to their flare ribbon areas. Pixels with positive polarity are colored blue, while pixels with negative polarity are colored red. The left panels show newly brightened pixels, while the right panels show the total accumulation flare ribbon pixels.

Here we provide an example of the flare ribbon evolution in our AIA image frame snapshot phases, demonstrated in **Figure 4**, where the newly brightened pixels that exceeds the threshold are displayed in the left panels. In comparison, the total newly brightened pixel accumulation from the previous images up to that frame is displayed on the right panels. The corresponding new brightened flare areas are plotted on the HMI LOS magnetogram in the bottom row. The blue pixels indicate positive magnetic polarity, while red pixels indicate negative polarity.

## 3 Results and Discussion

In this section, we will discuss our results obtained: the correlation between flare classes with flare duration (Section 3.1) and magnetic reconnection flux (Section 3.2). Both sections use the same list of flare events, they are studied separately for confined (without CME) and eruptive (associate with CME) flare events. Linear correlation and the correlation coefficient calculated in log-log space were obtained and histograms were plotted to compare and analyse the characteristics of the flares.

### 3.1 Flare Class vs. Flare Duration

*3.1.1 Results*

The NOAA GOES X-ray time series data for each flare event were retrieved to derive the peak intensity and FWHM duration. The plots of GOES X ray flux vs. time for each of the 33 events are visualised in **Appendix A** for reference, while their data are summarised in **Table 2**. In the table, the events are arranged by flare type (events no. 1–17 are confined, 18–33 are eruptive), then in ascending order of GOES classes, starting from M5 to higher X classes.

The X-ray peak intensities ($Wm^{-2}$) and FWHM duration in **Table 2** were used to plot **Figure 5** to find the Pearson's linear correlation between flare class and duration (FWHM) for the 33 events. The regression line in the left plot of **Figure 5** was done using standard linear correlation while the one on the right was done by taking the logarithm of both variables. The linear correlation graphs are displayed on two separate scales to better compare with earlier literature (see **Section 3.1.2**). In **Figure 5** (bottom), the flare classes plotted against duration were segregated by flare type to compare their individual linear correlation coefficient values (*r*).

**Table 2**: The GOES X-ray peak intensities, flare duration (FWHM), and total accumulated magnetic reconnection flux for each flare event. Events no. 1–17 are confined flares, while 18–33 are eruptive flares.

| No. | Event (Date) | GOES Class | X-Ray Peak Intensity ($10^{-4}$ W m$^{-2}$) | FWHM ($t_2 - t_1$, s) | Total Accumulated Magnetic Reconnection Flux ($10^{21}$ Mx) | | |
|---|---|---|---|---|---|---|---|
| | | | | | Positive Polarity | Negative Polarity | Mean |
| 1 | 3-Nov-13 | M5.0 | 0.52 | 272 | 1.29 | 0.78 | 1.03 |
| 2 | 4-Feb-14 | M5.2 | 0.52 | 508 | 3.56 | 2.37 | 2.97 |
| 3 | 4-Jul-12 | M5.3 | 0.55 | 240 | 5.37 | 1.63 | 3.50 |
| 4 | 24-Aug-15 | M5.6 | 0.59 | 166 | 0.56 | 1.37 | 0.96 |
| 5 | 10-May-12 | M5.7 | 0.60 | 381 | 4.82 | 2.19 | 3.51 |
| 6 | 5-Jul-12 | M6.1 | 0.64 | 328 | 4.88 | 0.90 | 2.89 |
| 7 | 13-Feb-11 | M6.6 | 0.67 | 717 | 3.33 | 3.80 | 3.56 |
| 8 | 18-Dec-14 | M6.9 | 0.70 | 1919 | 2.61 | 3.25 | 2.93 |
| 9 | 7-Jan-14 | M7.2 | 0.73 | 1542 | 2.52 | 3.54 | 3.03 |
| 10 | 28-Sep-15 | M7.6 | 0.77 | 430 | 2.02 | 0.55 | 1.28 |
| 11 | 22-Oct-14 | M8.7 | 0.88 | 2742 | 3.36 | 4.80 | 4.08 |
| 12 | 30-Jul-11 | M9.3 | 0.96 | 223 | 2.20 | 1.61 | 1.90 |
| 13 | 19-Oct-14 | X1.1 | 1.10 | 4121 | 4.17 | 2.01 | 3.09 |
| 14 | 9-Mar-11 | X1.5 | 1.58 | 459 | 5.13 | 6.72 | 5.93 |
| 15 | 22-Oct-14 | X1.6 | 1.67 | 2378 | 6.04 | 6.96 | 6.50 |
| 16 | 26-Oct-14 | X2.0 | 2.01 | 1939 | 6.74 | 5.15 | 5.95 |
| 17 | 24-Oct-14 | X3.1 | 3.20 | 3406 | 9.79 | 8.68 | 9.24 |
| 18 | 10-Mar-15 | M5.1 | 0.53 | 385 | 0.89 | 1.17 | 1.03 |
| 19 | 28-Sep-14 | M5.1 | 0.51 | 1903 | 0.09 | 0.30 | 0.20 |
| 20 | 6-Sep-11 | M5.3 | 0.54 | 1237 | 1.50 | 3.40 | 2.45 |
| 21 | 4-Dec-14 | M6.1 | 0.62 | 2353 | 2.75 | 2.14 | 2.44 |
| 22 | 9-Mar-12 | M6.3 | 0.64 | 2275 | 5.95 | 5.35 | 5.65 |
| 23 | 22-Jun-15 | M6.6 | 0.66 | 3262 | 4.51 | 5.86 | 5.19 |
| 24 | 8-Sep-11 | M6.7 | 0.68 | 700 | 1.52 | 4.35 | 2.94 |
| 25 | 7-Sep-17 | M7.3 | 1.01 | 123 | 0.44 | 2.64 | 1.54 |
| 26 | 25-Jun-15 | M7.9 | 0.80 | 3045 | 2.72 | 3.30 | 3.01 |
| 27 | 17-Dec-14 | M8.7 | 0.88 | 2650 | 1.95 | 3.70 | 2.82 |
| 28 | 4-Aug-11 | M9.3 | 0.94 | 633 | 2.49 | 2.92 | 2.71 |
| 29 | 8-Nov-13 | X1.1 | 1.22 | 229 | 0.65 | 3.36 | 2.01 |
| 30 | 12-Jul-12 | X1.4 | 1.42 | 3987 | 5.63 | 7.32 | 6.48 |
| 31 | 10-Sep-14 | X1.6 | 1.67 | 2879 | 5.18 | 5.45 | 5.32 |
| 32 | 11-Mar-15 | X2.2 | 2.22 | 614 | 2.46 | 3.05 | 2.76 |
| 33 | 7-Mar-12 | X5.4 | 5.44 | 1372 | 6.47 | 8.70 | 7.59 |

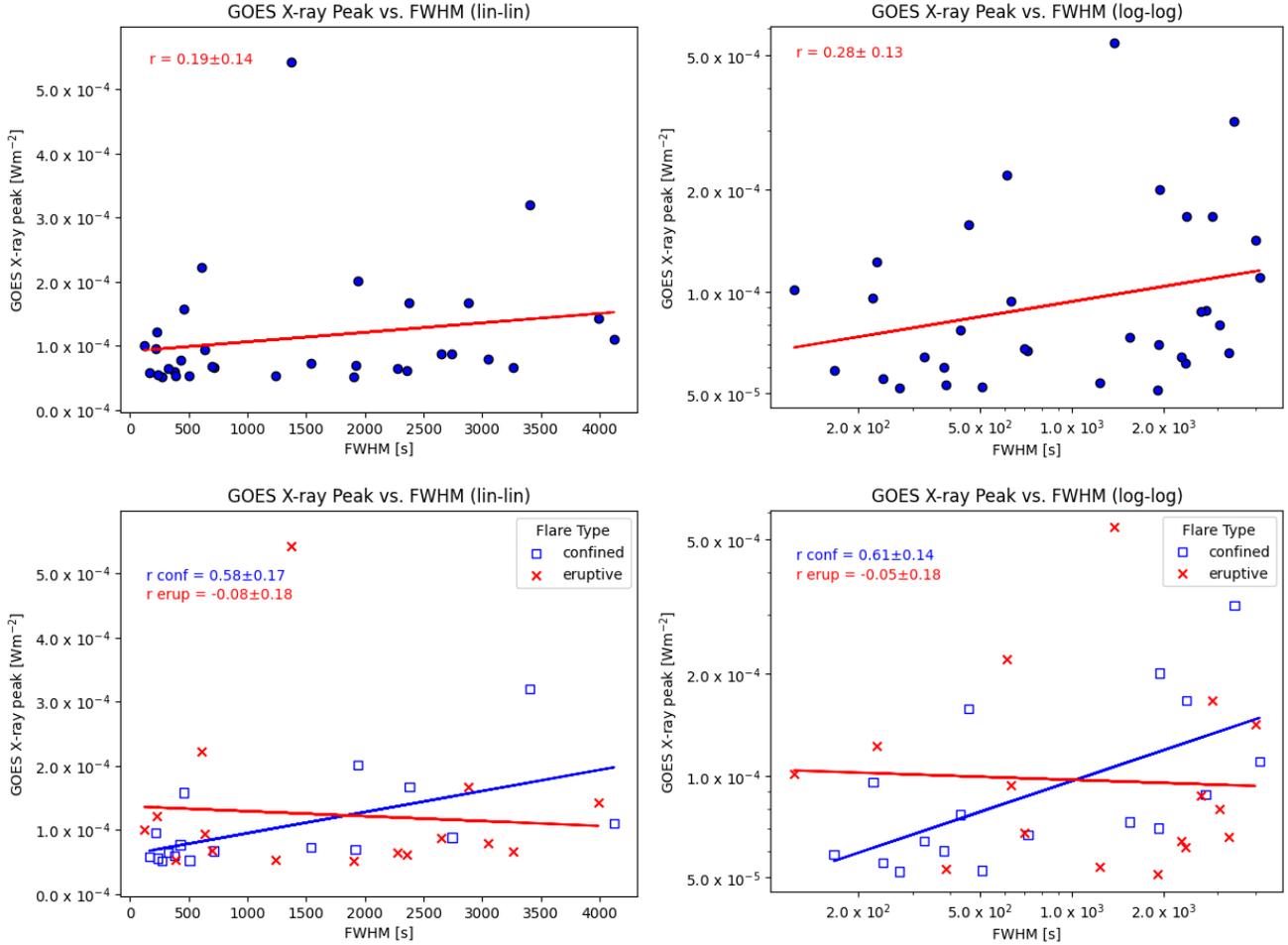

**Figure 5**: (top) GOES X-ray peak magnitude (flare class) vs. FWHM for all 33 flares events and (bottom) separated by flare type. The regression lines and correlation coefficients for confined ($r_{conf}$) and eruptive ($r_{erup}$), were determined on linear-linear (left) and log-log (right) scales, respectively.

### 3.1.2 Discussion

In general, the best linear correlation coefficient value ($r$) obtained when events of both flare types are used is $r = 0.28 \pm 0.13$, when the log of the GOES X-ray peak magnitude is plotted against the log of the flare duration. **Table 3** summarises the correlation coefficient values obtained in this work as compared to those in the literature shared by linear and log-log space.

When considering all flare types, the correlation coefficient value obtained here shows weak correlation, comparable with the value reported by Veronig et al. (2002), showing $r = 0.25$ for a log-log relationship. On the other hand, our linear coefficient value obtained using the linear scale shows a slightly higher value of $r = 0.19 \pm 0.14$, compared to $r = 0.09$ found by Reep and Knizhnik (2019), who noted that the duration is independent of flare class, other than believing that higher flare classes with larger energy should have a longer duration time.

**Table 3**: The Pearson correlation coefficient values (*r*) of flare class vs. flare duration in log-log and linear-linear scales.

| Authors | Correlation Coefficient, *r* (Flare Class vs. Flare Duration) | | | | | |
|---|---|---|---|---|---|---|
| | All Flares | | Confined | | Eruptive | |
| | log-log | lin-lin | log-log | lin-lin | log-log | lin-lin |
| Veronig et al. (2002) | 0.25 | - | - | - | - | - |
| Reep & Knizhnik (2019) | - | 0.09 | - | - | - | - |
| This Work | 0.28 | 0.19 | 0.61 | 0.58 | -0.05 | –0.08 |

In general, we identified a weak linear correlation between the GOES X-ray peak and flare duration prior to the segregation of flare types, agreeing with the results of Veronig et al. (2002) and Reep & Knizhnik (2019). However, further dividing the sample by flare type (as shown in **Figure 5**) reveals that confined flares have a higher linear correlations coefficient ($r = 0.58 \pm 0.17$) as compared to eruptive flares ($r = –0.08 \pm 0.18$), which showed no correlation. Similarly, the linear correlation coefficients for the confined flares and eruptive flares were also calculated in log-log space, with ($r = 0.61 \pm 0.14$) and ($r = –0.05 \pm 0.18$), respectively. This result agrees with the work of Kay et al. (2003), which states a correlation between the GOES X-ray peak and flare duration trend observed in the confined flares, while more scattered with no correlation in the eruptive flares. We note that Kay et al. (2003) did not share their correlation coefficient values but shared that there is a general correlation between the duration and flare class for all the events, with high peak intensity events generally being of longer duration, which is consistent with the work of Veronig et al. (2002).

In this study, the bootstrap method has been used to estimate the standard errors of the correlation coefficients. The confined flare correlation coefficient ($r = 0.58 \pm 0.17$) is not highly precise, still provides a useful estimate of the strength between the GOES X-ray peak and flare duration. However, the range of uncertainty is relatively large for all flares and eruptive flare types, indicating a high uncertainty. A weak relationship between flare class and flare duration among the eruptive flares can make it more challenging to estimate the true correlation coefficient accurately, resulting in a larger uncertainty range. The low value of the correlation coefficient, along with a small sample size, could contribute to the larger uncertainty level.

Here we emphasise that the difference in sample size is an important statistical aspect that needs to be considered despite having our correlation coefficients for GOES X-ray peak and flare duration almost comparable to those of and slightly higher than those of Reep & Knizhnik (2019). This is because our sample's flare classes were taken from a significantly different energy level. Previous studies had a skewed sample size towards certain lower flare classes (B and C classes), which may not be representative of the higher flare class characteristics, especially X classes. For example, Reep and Knizhnik (2019) selected 2956 events from April 2010 to April 2016, which comprised of 91.3% C class, 8.2% M class and 0.5% X class flares, respectively. Veronig et al.(2002) studied the temporal characteristics of 26745 events between January 1998 to December 2000, consisting of 33.1% B class, 61.7% C class, 5.0% M class and 0.2% X class flares, respectively. Noticeably, both their sample sizes are highly skewed towards C flare classes compared to X flare classes population, which are between 0.2% to 0.5% only.

The lower number of X class flares is due to the rare occurrence of the higher flare classes. In our study, we selected 33 events between February 2011 and September 2017, with a focus on flare classes above M5, consisting of 70% M class and 30% X class flares, giving a statistically comparable sample size between each class. B and C class flares were not included in our sample size due to their insignificant impact on Earth and is close to the high background flux (radiation emitted during no solar flare). Other than the correlation trend and sample size, we have also plotted a bar chart as shown in

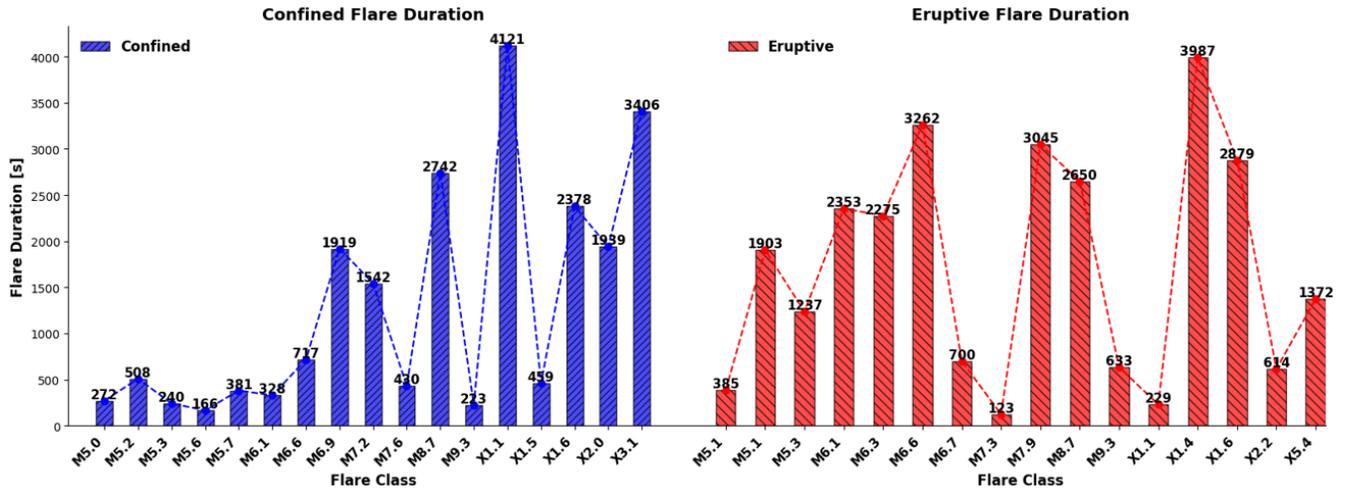

**Figure 6:** Flare duration display by flares classes, segregated by flare type confined (left) and eruptive (right) events.

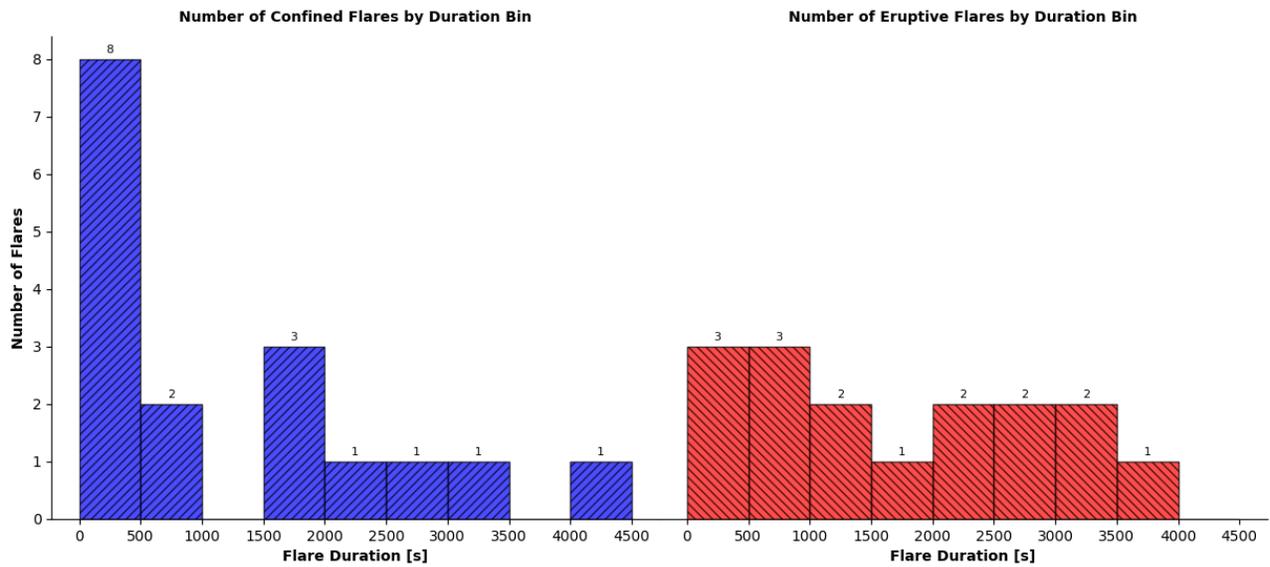

**Figure 7**: The number of occurrences of flares by flare duration is shown in a histogram, segregated by confined (left) and eruptive (right) flare types.

**Figure 6** to visually assess if there are any significant differences or similarities in the flare duration display by flare class, separately studied according to their flare type.

The visual comparison between confined and eruptive flares of the bar chart in **Figure 6** shows that the duration of confined flares majority is shorter than the eruptive flares of the same M-class. This is reflected in the histogram in **Figure 7**, where 8 confined flare events occurred in the first bin of the duration within 500 s, compared to 3 eruptive events in the same duration bin.

**Table 4** shows the minimum, maximum and average flare durations for the whole flare data set as well as separately by flare class and type (eruptive / confined). A general observation here is that both confined and eruptive flares have almost the same minimum and maximum duration, between 120 to 4000 seconds, reflecting a balanced sample size in the current study.

Table 4: Flare duration, separated by flare types and GOES classes.

|  | Flare Duration (s) | | | M Class | X Class |
|---|---|---|---|---|---|
|  | All Classes | | | | |
|  | Minimum | Maximum | Average | Average | Average |
| **All Flares** | 123 | 4121 | 1498 | 1219 | 2138 |
| **Confined** | 166 | 4121 | 1281 | 789 | 2460 |
| **Eruptive** | 123 | 3987 | 1728 | 1688 | 1817 |

Another observation is that the average duration increased from M to X class flares by a factor of 3 for confined flares (789 s → 2460 s), but insignificantly for eruptive flares (1688 s → 1817 s). It is also notable that the average duration of confined M class flares (789 s) is shorter, about half of the eruptive M class flares (1688 s).

Generally, confined flares have shown a higher correlation with flare duration (**Fig. 5**). This trend is in agreement with the expectation of Reep and Knizhnik (2019) that higher class flares with higher magnitude energies are expected to have a longer flare duration, even though it was not clearly distinguished by flare type. Similar to our results, Kay et al. (2003) observed a trend where the correlation between GOES peak intensity and flare duration becomes more significant when only considering events that are not associated with CME. The visual comparison (**Fig. 6**) by flare type shows that the duration of eruptive flares is typically longer than that of confined flares of the same M-class events where their average duration of the confined M class flares is about half of the eruptive flares (**Table. 4**). It is also, reflected in the histogram (**Fig. 7**) where almost half of the confined events sample size falls in the shortest duration bin (500 s) compared to eruptive events.

This result suggests that confined M class flares generally require a shorter time for a chromosphere evaporation process to emit SXR emission, which agrees with the findings of Kay et al. (2003) that showed a tendency for confined flares to heat the plasma to a given temperature faster than eruptive flares. The energy-dependent process of chromosphere evaporation in eruptive flares, which is associated with the CME, takes a longer time to accomplish. The CME involves significant restructuring of the magnetic flux in the corona, which may cause the timescale for energy release to be longer in eruptive flares than confined flares (Kay et al., 2003). This corresponds to the findings of Green et al. (2001), who noticed that less energy goes into heating the plasma when associated with CME, with the more energy potentially going into the ejection of the material in the CME.

The heating of the plasma towards chromosphere evaporation relates to the rise phase of the flare. Another important parameter for characterising the temporal evolution of solar flares is the decay phase, where the FWHM measures both the rising and falling time of the light curve. The decay phase of a solar flare is related to the cooling of the plasma in the flare loops. This cooling time depends on the length of the loop, with a cooling time scales as $L^{5/6}$, where $L$ is the length of the loop (Cargill et al., 1995). In a related aspect, a clear linear correlation has been found between the separation of ribbon centroids against the FWHM and e-folding decay times of GOES light curves, which has been studied for flare class M5 and larger (Reep & Toriumi, 2017).

## 3.2 Flare Class vs. Total Magnetic Reconnection Flux

### 3.2.1 Results

The total magnetic reconnection flux for each flare was obtained after processing the images of 1600 Å AIA and HMI line-of-sight magnetic field maps using an algorithm developed using SUNPY. The total magnetic reconnection flux trendlines of all 33 events are displayed in **Appendix A** and tabulated in **Table 2**. The accumulated magnetic reconnection fluxes of positive and negative polarity are averaged and plotted against the GOES X-ray peak magnitude, as displayed in **Figure 8,** showing the correlation comparison before and after segregation by flare types.

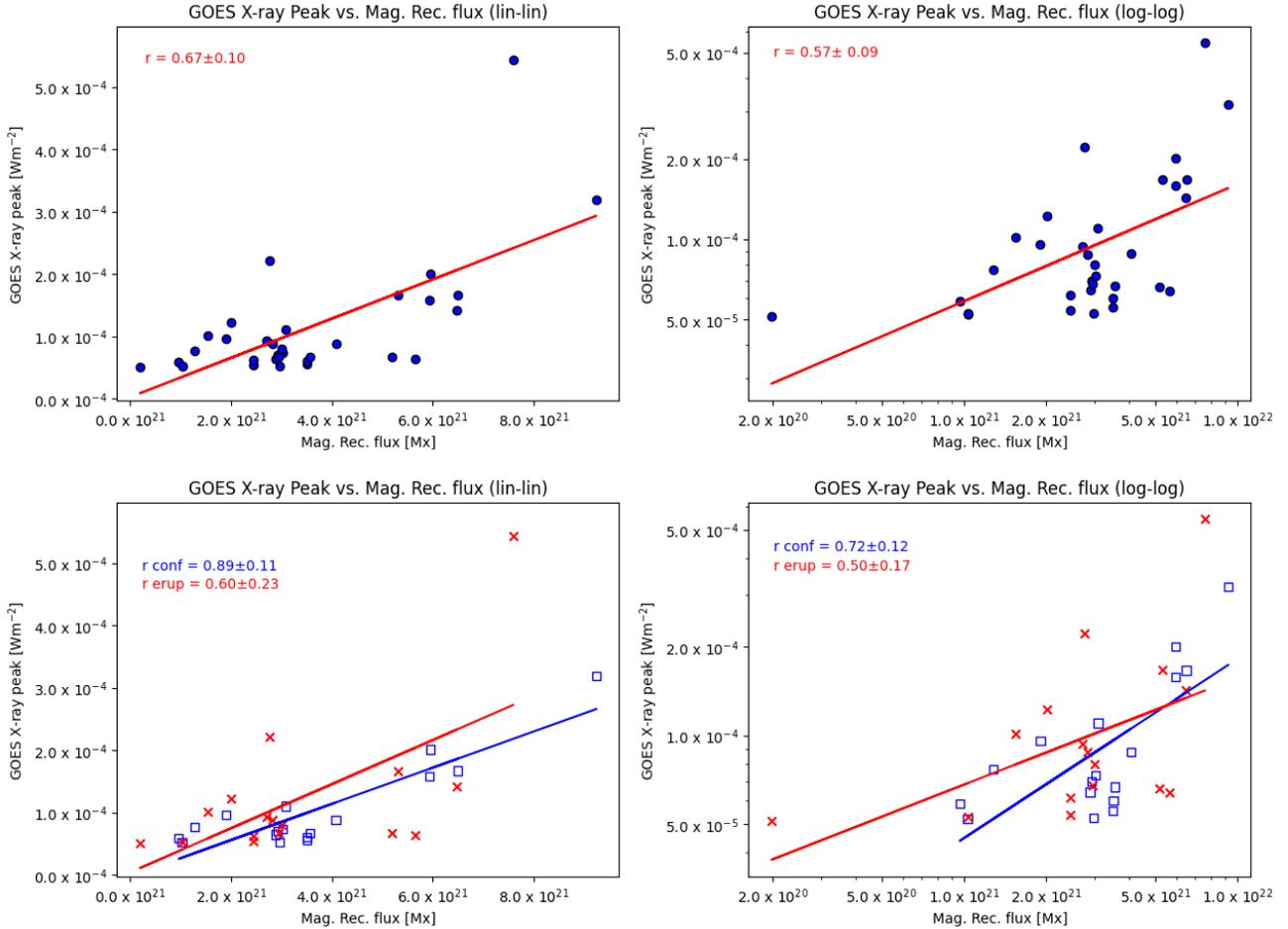

**Figure 8**: GOES X-ray peak magnitude (flare class) vs. magnetic reconnection flux for all 33 flare events (top) and those separated by flare type (bottom). The regression lines and correlation coefficients for confined ($r_{conf}$) and eruptive ($r_{erup}$) were determined on linear-linear (left) and log-log (right) scales, respectively.

### 3.2.2 Discussion

In general, when not separated by flare type, plotting the GOES X-ray peak magnitude vs. magnetic reconnection flux gives a linear correlation coefficient of $r = 0.67 \pm 0.11$ (**Figure 8**). Previous similar studies by Veronig & Polanec (2015) and Tschernitz et al. (2018) give correlation coefficients of $r = 0.78$ and $r = 0.92$, respectively. **Table 5** shows the results of this work as compared to others in literature.

Table 5: The Pearson correlation coefficient values (*r*) of flare class vs. total magnetic reconnection flux, derived in log-log and standard (lin-lin) scales, comparing the results of our work with the literature.

| Authors | | Correlation Coefficient, *r* (Flare Class vs. Magnetic Reconnection Flux) | | | | | |
|---|---|---|---|---|---|---|---|
| | | All Flares | | Confined | | Eruptive | |
| | | log-log | Standard | log-log | Standard | log-log | Standard |
| **Veronig & Polanec (2015)** | All Classes | - | 0.78 | - | - | - | - |
| **Tschernitz et al. (2018)** | All Classes | 0.92 | 0.78 | 0.90 | 0.93 | 0.94 | 0.78 |
| **This Work** | All Classes | 0.57 | 0.67 | 0.72 | 0.89 | 0.50 | 0.60 |

Veronig & Polanec (2015) compared their single confined X1.6 flare with the values obtained from 27 eruptive flares retrieved from archive data. On the other hand, Tschernitz et al. (2018) used a sample size of 51 flares which consists of 14% B class, 29% C class, 35% M class and 22% X class flares, respectively, with events from Jun 2000 to Jun 2015. The events were selected to cover representative samples of both eruptive and confined flares in all GOES classes. Their published correlation coefficient for flare class vs. total magnetic reconnection flux was derived from a log-log correlation and included extreme flares with classes X10 and X17. As Miklenic et al. (2009) previously noted, when a few extreme events are included, the correlation between the flare magnetic reconnection flux and the GOES class will be significantly higher. Therefore, we are unable to make a direct comparison with their results as we do not include extreme events.

We analysed the relationship between reconnection flux and the GOES class using both a log-log transformation and a linear-linear plot. The correlation coefficient in the log-log transformation space is $0.57 \pm 0.09$, while in the linear plot, it is $0.67 \pm 0.10$. Notably, there are no extreme events included in the analysis.

When examining the 51-event data of Tschernitz et al. (2018), we observed a change in their linear correlation coefficient for the total magnetic reconnection flux vs. the GOES class. Upon converting their data from a log-log transformation to a linear-linear plot and excluding extreme events, their correlation coefficient shifted from 0.92 to 0.78. Similarly, Veronig & Polanec (2015) have shared their standard linear correlation coefficient after excluding the extreme events, which saw a reduction of the coefficient from 0.78 to 0.64.

Comparing these correlation coefficient values, we found that our coefficient of 0.67 in the linear plot is comparable to both Veronig & Polanec (2015) value of 0.64 and Tschernitz et al. (2018) adjusted value of 0.78 (after excluding extreme events), before segregating by flare type. These findings suggest there is a consistent and significant correlation between reconnection flux and the GOES class.

In **Figure 8**, we presented a plot of the GOES X-ray peak magnitude against the total magnetic reconnection flux, with the data segregated by flare type. We observed that confined flares exhibit a higher linear correlation coefficient of $0.89 \pm 0.11$, while eruptive flares show a slightly lower coefficient of $0.60 \pm 0.23$. Similarly, when applying the log-log transformation, the correlation coefficients for confined and eruptive flares are $0.72 \pm 0.12$ and $0.50 \pm 0.17$, respectively. Our results reveal that confined flares have a higher linear correlation coefficient than eruptive flares, which was not the case observed by Tschernitz et al. (2018). However, when the data of Tschernitz et al. (2018) were converted to standard linear correlation and extreme flares were omitted, the correlation coefficients for flare class vs. total magnetic reconnection flux for confined and eruptive flares varied from 0.90 to 0.93 and 0.94 to 0.78, respectively. Though both continue to have a high correlation coefficient, the confined coefficient value has risen above eruptive flares, consistent with our trend.

To summarise, our analysis demonstrates distinct correlation patterns between the GOES X-ray peak magnitude and the total magnetic reconnection flux for confined and eruptive flares. The difference in correlation coefficients by flare type highlights the importance of considering these distinctions in the interpretation of solar flare data.

The bar chart in **Figure 9** is plotted to visually determine whether there are any notable differences or similarities in the magnetic reconnection flux display by flare class, separately studied according to their flare type. We could not significantly visualise any major differences, except for a minor difference where a couple of confined X-class flares stand higher than their counterparts in the eruptive events. Similarly, histogram **Figure 10** shows an overall trend comparable, with the most occurrence of total magnetic reconnection flux falling within $2.0 \times 10^{21}$ Mx to $4.0 \times 10^{21}$ Mx. This reflects the equal strength of the confined and eruptive sample sizes in our study.

Based on the data presented in **Table 6**, it can be observed that confined M-class solar flares exhibit an average total magnetic reconnection flux of $2.64 \times 10^{21}$ Mx, a value quite comparable to their eruptive counterparts at $2.73 \times 10^{21}$ Mx. However, the scenario changes for X-class flares, where confined flares demonstrate a notably higher average total magnetic reconnection flux of $6.14 \times 10^{21}$ Mx, in contrast to their eruptive counterparts which average at $4.83 \times 10^{21}$ Mx. Notably, our findings highlight distinct energy propagation characteristics between confined and eruptive flares, with the disparities being most prominent in the context of X-class flares.

Our data indicated that confined flares and eruptive flares have different energy propagation characteristics, and these differences are most pronounced in the case of X-class flares. Confined flares are characterised by a release of energy that is entirely absorbed by the chromosphere through the process of chromospheric evaporation, resulting in intense SXR emission. On the other hand, eruptive flares are accompanied by coronal mass ejections (CMEs), where part of the magnetic reconnection energy is converted to kinetic energy and escapes as CME. Statistically, the CME kinetic energy is a larger contribution than the energy in the associated flare (Emslie et al., 2005).

Table 6: Magnetic reconnection flux, separated by flare types and GOES classes.

|  | Magnetic Reconnection Flux ($10^{21}$ Mx) | | | | |
|---|---|---|---|---|---|
|  | All Classes | | | M Class | X Class |
|  | Minimum | Maximum | Average | Average | Average |
| **All Flares** | 0.20 | 9.24 | 3.53 | 2.68 | 5.48 |
| **Confined** | 0.96 | 9.24 | 3.67 | 2.64 | 6.14 |
| **Eruptive** | 0.20 | 7.59 | 3.38 | 2.73 | 4.83 |

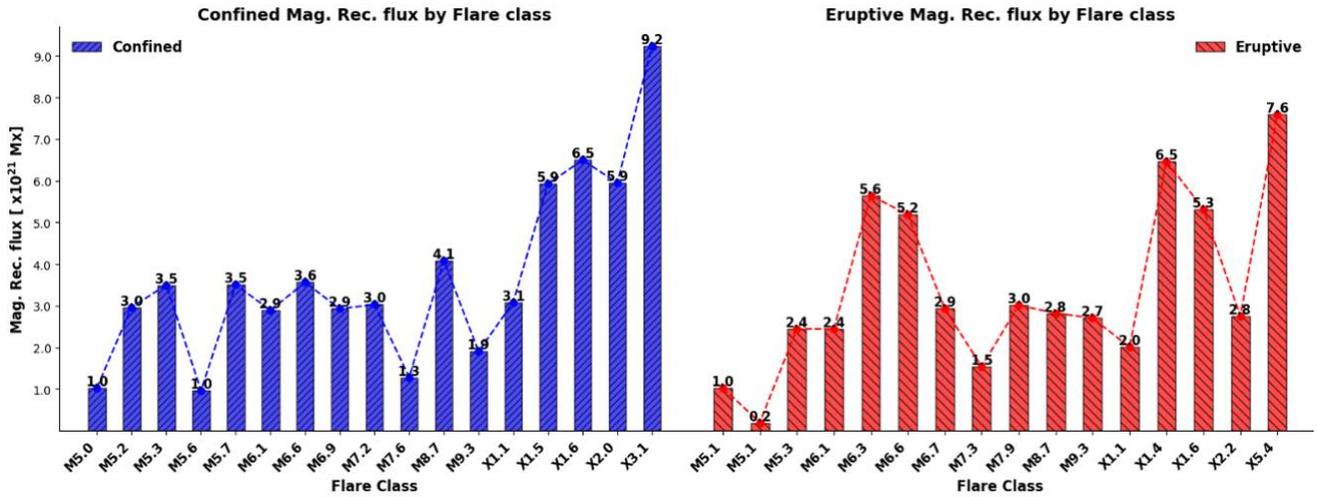

**Figure 9**: Flare magnetic reconnection flux display by flares classes, segregated by flare type confined (left) and eruptive (right) events.

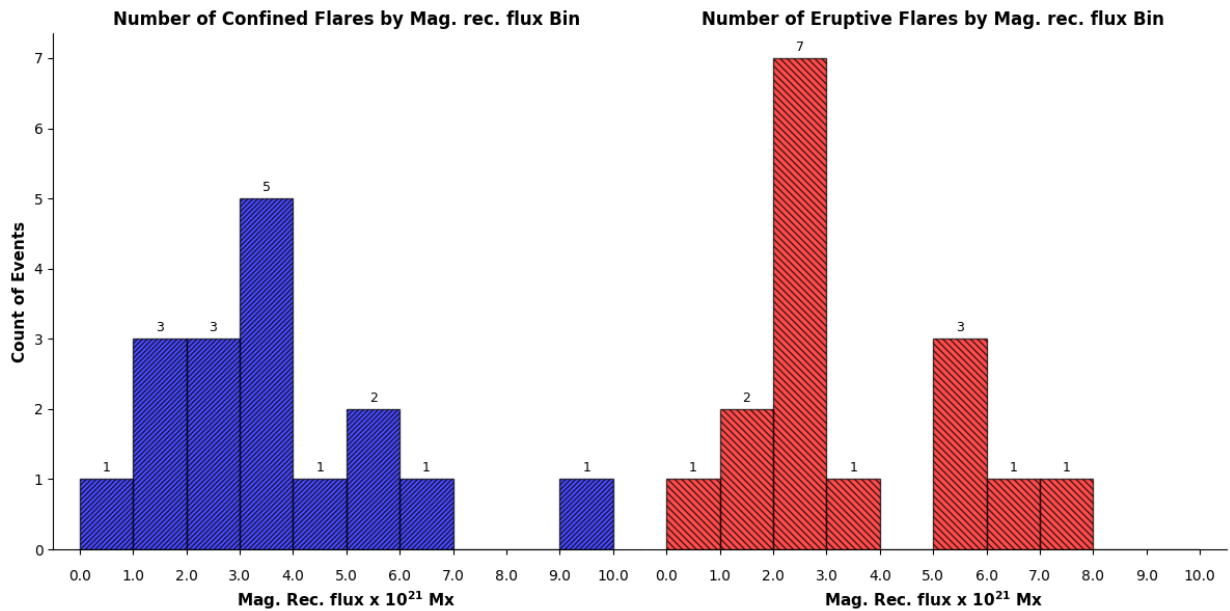

**Figure 10:** Histogram of the number of occurrences of flares by magnetic reconnection flux, segregated by confined (left) and eruptive (right) flare types.

## 4 Conclusion

Magnetic reconnection flux and flare duration (FWHM) were studied in 33 flare events with heliographic longitudes less than 45°. The event dates cover February 2011 to September 2017, representing from beginning to end of Solar Cycle 24. The flare X-ray peak magnitudes corresponding to GOES classes M5 – X5 represented by 17 confined and 16 eruptive flares were studied. In this work, we conclude as follows:

1. *Linear correlation is found between flare class and duration for confined events, but not for eruptive events.*
   We found that the linear correlation between flare class against duration (FWHM) is weak ($r = 0.19 \pm 0.14$) prior to the segregation by flare types, after segregation it is found that confined flares have a higher linear correlation ($r = 0.58 \pm 0.17$), while close-to-no correlation ($r = -0.08 \pm 0.18$) is observed for eruptive flares. Their correlation coefficient values in the log-log space are $r = 0.28 \pm 0.13$ prior to segregation, while after segregation by flare event type, they are $r = 0.61 \pm 0.14$ for confined events and $r = -0.05 \pm 0.18$ for eruptive events, respectively. The current study correlations support those previous findings by Veronig et al. (2002), Reep & Knizhnik (2019), and Kay et al. (2003). Therefore, we believe that the general assumption that duration is independent of flare class without flare event segregation is inconclusive, and further investigation should be done with a larger sample size.

2. *The confined M-class flare durations are about half of the durations of eruptive flares.*
   Our results have shown that confined M class average flare durations (789 s) are less than half of those of eruptive flares (1688 s), and the histogram also reflected that half of the confined events occurred within the shortest duration bin of 500 s. This indicates that in confined M class flares without CME, the plasma is heated faster, which agrees with the findings of Kay et al. (2003). Confined flares do not result in a coronal mass ejection (CME), therefore preventing the released energy and heat from escaping into interplanetary space. Instead, it directs the released energy and heat towards the chromosphere. This results in a build-up of energy and pressure within the limited region, leading to rapid heating of the chromosphere. This rapid heating of the chromosphere during a confined flare can cause it to rapidly expand and evaporate, leading to a phenomenon known as chromosphere evaporation (Tschernitz et al., 2018) that emits SXR emissions. On the other hand, eruptive flares require a longer duration to complete the energy-dependent chromosphere evaporation process, where the CME associated with the eruption is responsible for the delay in energy propagation since it involves a massive restructuring of the corona's magnetic flux (Kay et al., 2003). This is in line with the observations made by Green et al. (2001), who determined that less energy goes into heating the plasma when associated with a CME, with more energy potentially going into the ejection of the material in the CME.

3. *The linear correlation between flare class and total magnetic reconnection flux for confined events is greater than that for eruptive events.*
   Prior to the segregation of flare types, we observed a significant linear correlation ($r = 0.67 \pm 0.10$) between flare class and total magnetic reconnection flux. Following the segregation of flare types, our results indicate that confined flares have a greater linear correlation ($r = 0.89 \pm 0.11$) than eruptive flares ($r = 0.60 \pm 0.23$). Their correlation coefficient values in the log-log space prior to segregation ($r = 0.57 \pm 0.09$) and after segregation by flare type, confined events ($r = 0.72 \pm 0.12$) and eruptive events ($r = 0.50 \pm 0.17$) obtained for better comparison. Our linear correlation between flare class and total magnetic reconnection flux findings are consistent with previous work by Tschernitz et al. (2018) and Veronig & Polanec (2015).

The linear correlation between the flare type and the total magnetic reconnection flux is stronger for confined events compared to eruptive events due to differences in the physical mechanisms involved. Confined flares involve the release of energy within the corona without any substantial eruption of matter. This results in a uniform and controlled release of magnetic reconnection energy, which is confined within the coronal loops, leading to a more predictable connection between the flare class and the total magnetic reconnection flux. On the other hand, eruptive flares are more complicated and include significant material eruptions, such as the formation of a coronal mass ejection (CME). As a result, some of the released magnetic energy is transformed into kinetic energy and is expelled as a CME. This making the connection between flare class and total magnetic reconnection flux less straightforward and weaker than in the case of confined flares. Therefore, the different mechanisms of energy release in confined and eruptive flares explain for the different linear correlation between flare class and total magnetic reconnection flux for the two types of events.

## 5 Future Work

In this work, we studied the correlation between magnetic reconnection flux and flare duration against flare class. In future work, we would like to conduct a regression analysis of total magnetic reconnection and flare duration in relation to the distance from the active region's magnetic core. Wang and Zhang (2007) investigated eight X-class flares and discovered that confined flares occur closer to the active region, whereas eruptive flares occur near the active region's edge. Baumgartner et al. (2018) discovered similar results after doing a statistical study of 44 flares. Thus, a regression analysis of total magnetic reconnection flux and flare duration vs. distance for both types of flares could enable the variables to be used to make prediction for both variables.

**Acknowledgements**   The author would like to express their gratitude to Prof. Dr. Astrid Veronig for the helpful guidance. This research has made use of SAOImageDS9, developed by Smithsonian Astrophysical Observatory. The SDO data used are courtesy of NASA/SDO and the AIA and HMI science teams. The authors appreciate NOAA for making the GOES SXR archive data available.

**Funding Note**       JYHS acknowledges financial support from the Fundamental Research Grant Scheme (FRGS) by the Malaysian Ministry of Higher Education with code FRGS/1/2023/STG07/USM/02/14. KB acknowledges support by the Malaysian Ministry of Higher Education through the MyPhD programme with account number KPT(B) 730410075801.

## Declarations

**Declaration of competing interest**      The authors declare that there are no conflicts of interest.

# Appendix A: GOES X-Ray Flux Plots and Magnetic Reconnection Flux Accumulation Trend

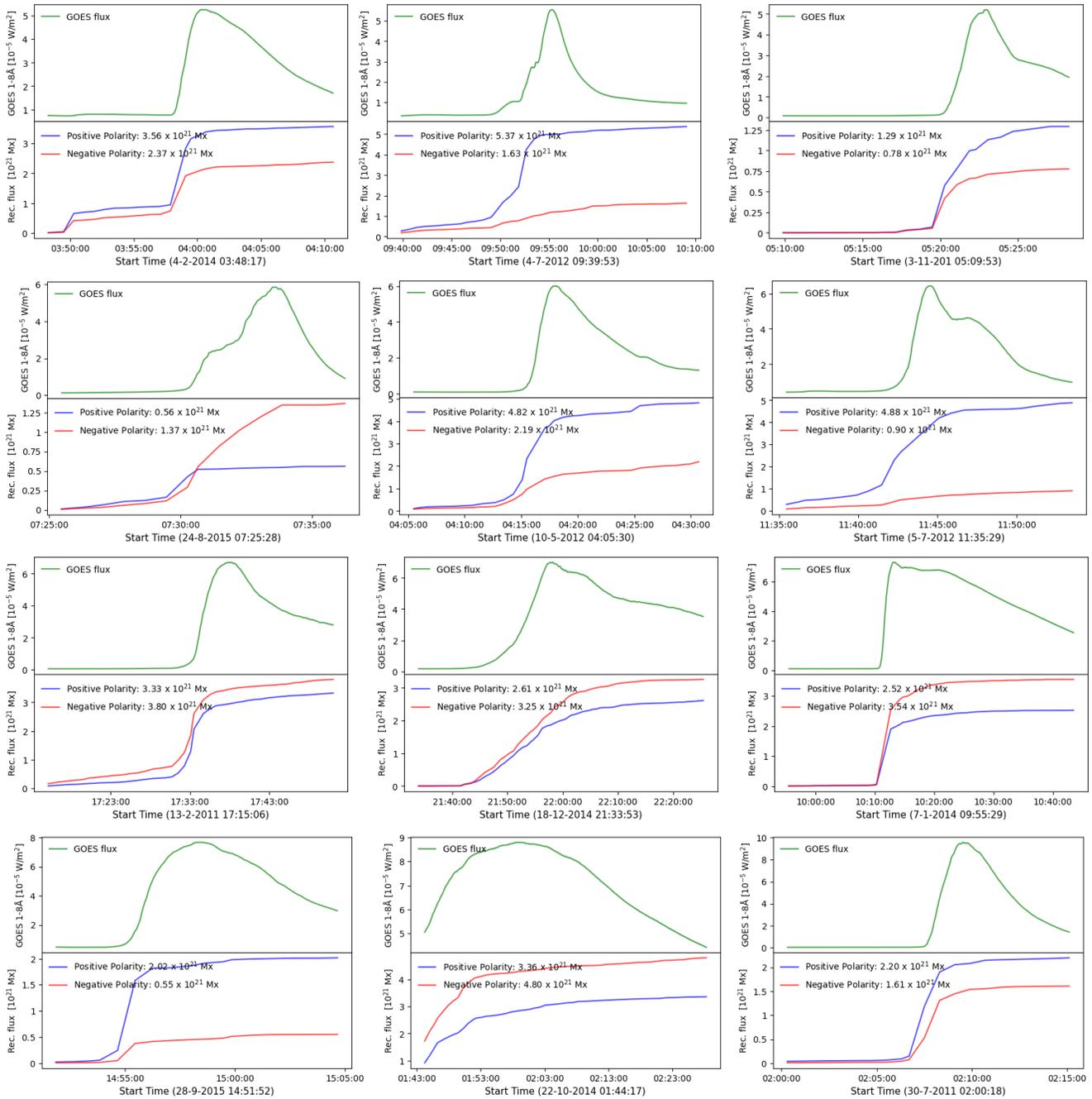

**Figure A1**: Display of the GOES X-Ray fluxes and magnetic reconnection flux trends for each event used in this study.

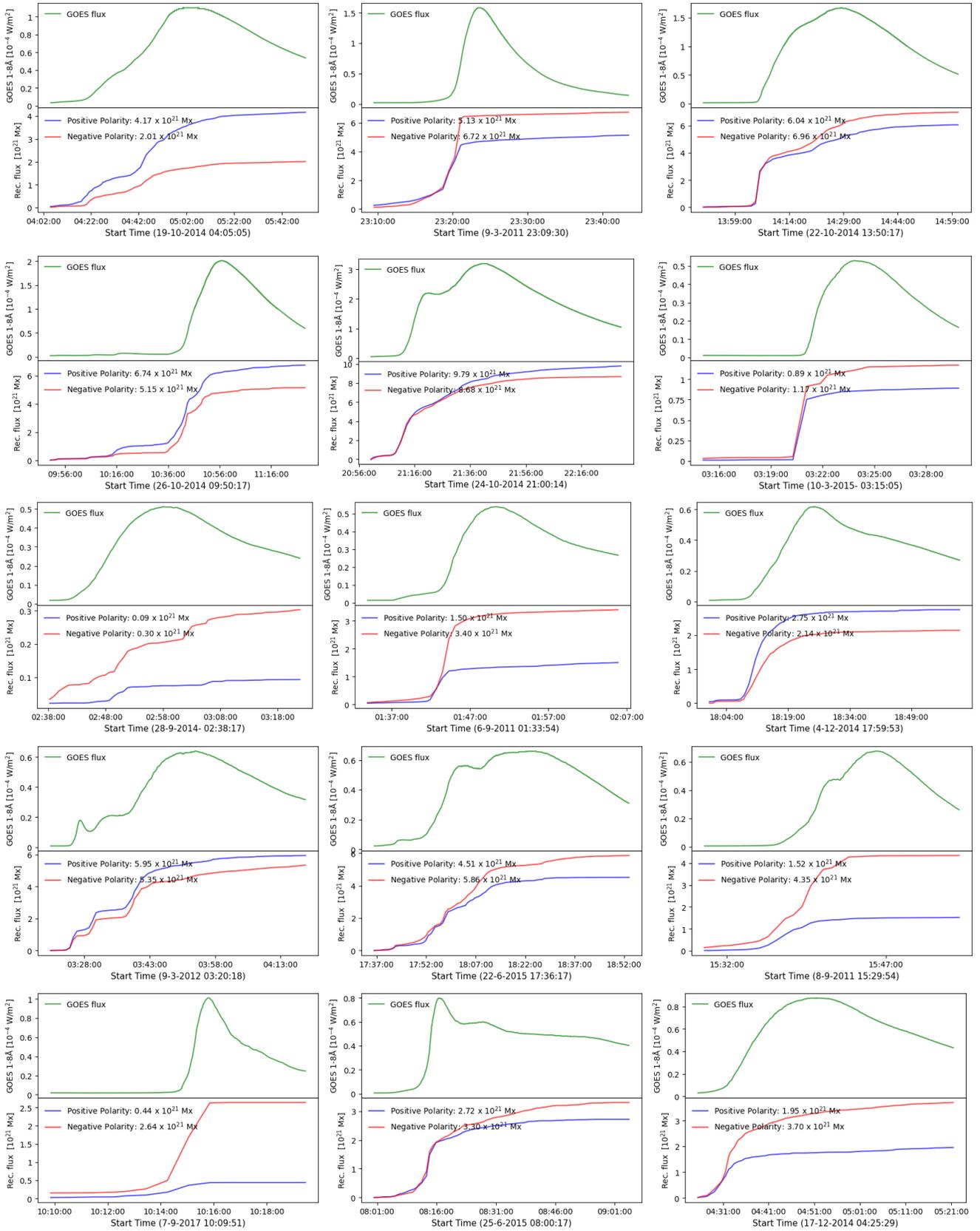

**Figure A1** (*cont'd*): Display of the GOES X-Ray fluxes and magnetic reconnection flux trends for each event used in this study.

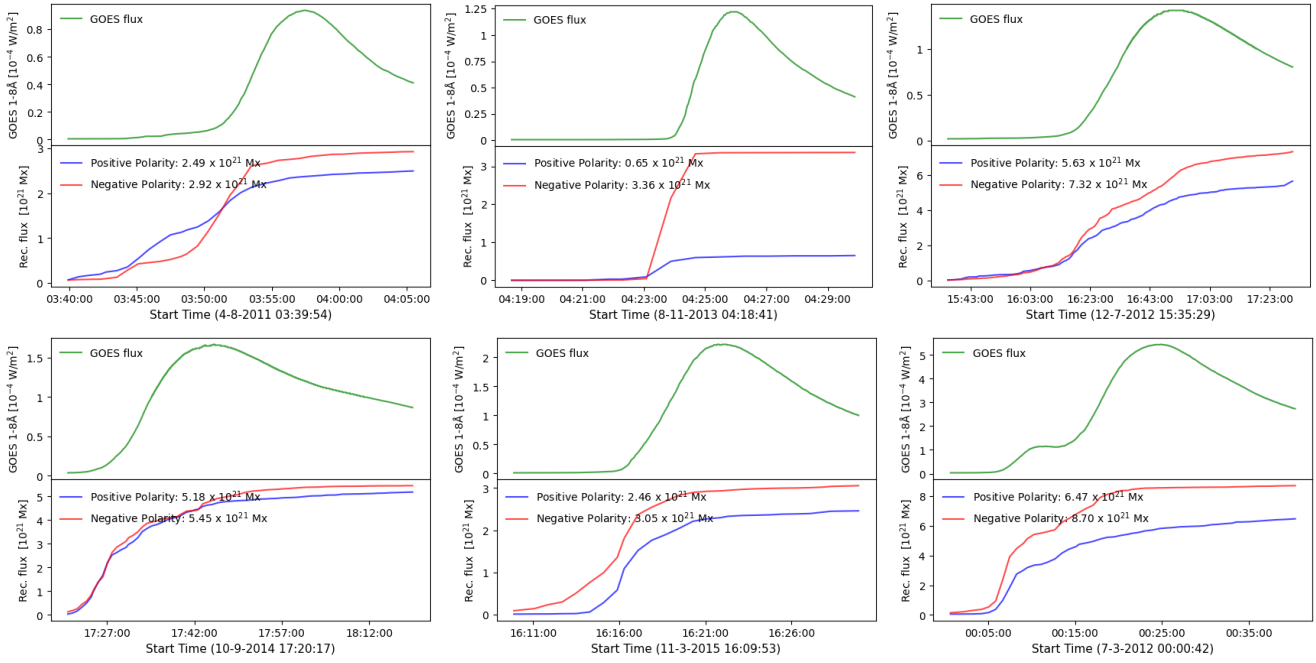

**Figure A1** (*cont'd*): Display of the GOES X-Ray fluxes and magnetic reconnection flux trends for each event used in this study.